\documentclass[12pt]{article}
\pdfoutput=1
\usepackage{amsmath}
\usepackage{amssymb}
\usepackage{amsthm}
\usepackage{psfrag}
\usepackage{graphicx}
\usepackage{hyperref}

\newcommand{\be}{\begin{equation}}
\newcommand{\ee}{\end{equation}}
\newcommand{\beqa}{\begin{eqnarray}}
\newcommand{\eeqa}{\end{eqnarray}}

\newcommand{\pd}{\partial}
\newcommand\m{\mu}
\newcommand\D{\Delta}
\newcommand\n{\nu}
\renewcommand\r{\rho}
\newcommand\s{\sigma}
\renewcommand\a{\alpha}
\renewcommand\b{\beta}
\renewcommand\l{\lambda}
\newcommand{\T}{\Theta}

\newcommand\vk{\varkappa}

\def\e{{\rm e}}
\def\d{\partial}
\newcommand{\bseq}{\begin{subequations}}
\newcommand{\eseq}{\end{subequations}}

\renewcommand{\ln}{\mathop{\rm ln}\nolimits}

\renewcommand{\Re}{\mathop{\rm Re}\nolimits}

\newcommand{\di}{\mathrm d}

\title{
\begin{center}
\sc{\huge UV-extending Ghost Inflation}
\date{}
\author{ 
Mikhail M. Ivanov$^{a,b,c}$\footnote{mm.ivanov@physics.msu.ru}~ 
and Sergey Sibiryakov$^{d,e,b}$\footnote{sergey.sibiryakov@cern.ch}
\vspace{.2cm}\\
\small\llap{$^a$}\it Faculty of Physics, Moscow State University,
\small \it Vorobjevy Gory, 119991 Moscow, Russia\\
\small\llap{$^b$}\it
Institute for Nuclear Research of the
Russian Academy of Sciences, \\ 
\small \it  60th October Anniversary Prospect, 7a, 117312
Moscow, Russia\\
\normalsize\llap{$^c$}
 \small \it Sternberg Astronomical Institute, Moscow State University,\\
 \normalsize\it Universitetsky prospect, 13, 119992 Moscow, Russia \\
\small\llap{$^d$}\it
Theory Group, Physics Department, CERN, CH-1211 Geneva 23,
Switzerland\\
\small\llap{$^e$}\it
FSB/ITP/LPPC, \'Ecole Polytechnique F\'ed\'erale de Lausanne,
\small \it CH-1015 Lausanne, Switzerland 
}
\end{center}
}

\begin{document}

\maketitle

\vspace{-12.5cm}
\begin{flushright}
CERN-PH-TH/2014-030 \\
INR-TH/2014-003
\end{flushright}

\vspace{9.5cm}

\begin{abstract}
We present a setup that provides a partial UV-completion of the
ghost inflation model up to a scale which can be almost as high
as the Planck mass. This is achieved by coupling the inflaton to the
Lorentz-violating sector described by the Einstein-aether theory or
its khronometric version. Compared to previous works on ghost
inflation our setup allows to 
go beyond the study of small
perturbations and include the background dynamics 
in a unified framework. In the specific regime when the expansion of the Universe is dominated by the kinetic energy of the inflaton we find that the model predicts rather high tensor-to-scalar ratio $r \sim 0.02\div 0.2$ and non-Gaussianity of equilateral type with $f_{NL}$ in the range from $-50$ to $-5$.

\end{abstract}

\newpage
\section{Introduction}
\label{sec:intro}

All structures in the observed Universe, such as galaxy clusters, galaxies
and stars are 
believed to arise from tiny quantum fluctuations, amplified during 
a primordial stage in the expansion of the Universe.
The most successful model of this stage is the inflationary theory which 
fits very well all current cosmological data.
Generally, inflation is supposed to occur at very high energies
(probably a few orders below the Planck mass), where the effects of new unknown 
physics may show up. Thus the inflationary 
period provides a unique opportunity to probe
these scales through cosmological observations. 
An intriguing possibility is that the known space-time symmetries
become invalid at the inflationary energies. One of such symmetries
is Lorentz invariance (LI) which is at the basis of the highly
successful Standard Model of particle physics and General Relativity
(GR). However, beautiful as it is, GR suffers from the problem of
non-renormalizability precluding it from being a consistent quantum
theory at energies above the Planck scale. It has been suggested by
Ho\v rava \cite{Horava:2009uw} that the situation can be improved by
allowing LI to 
be violated at high energy. A consistent implementation of this
proposal \cite{Blas:2009qj} involves additional light degrees of
freedom in the gravitational sector that introduce departures from LI
even at energies well below Planckian. The description at such
energies is provided by the so-called ``khronometric'' model
\cite{Blas:2010hb}, which can be considered as a variant of the
phenomenological ``Einstein-aether'' theory 
\cite{Jacobson:2000xp,Jacobson:2008aj} for the study of
Lorentz-violating (LV) effects in gravity (see
\cite{Jacobson:2010mx,Jacobson:2013xta} for the precise relationship
between the two models). The existing astrophysical and cosmological
data significantly 
constrain the parameters of the model 
\cite{Elliott:2005va,
Blas:2010hb,Blas:2011zd,Audren:2013dwa,Shao:2013wga,Yagi:2013qpa,
Yagi:2013ava}, but still leave open a theoretically motivated 
portion of the
parameter space. 
It is worth mentioning that any application of these ideas to 
realistic model building must include a mechanism that would prevent 
significant percolation of LI breaking from gravity into the Standard
Model sector where LI has been tested with an outstanding precision;
several options have been discussed in 
\cite{GrootNibbelink:2004za,Pospelov:2010mp,Pujolas:2011sk,Bednik:2013nxa}.

LV in gravity, though tightly
constrained at low energies accessible to current experiments, may be
significantly stronger during inflation.
An interesting alternative to the standard slow-roll
inflation involving LV in gravity is provided by the ghost inflation model
\cite{ArkaniHamed:2003uz} and its ``tilted'' extension
\cite{Senatore:2004rj}. The important feature of these models is the
modified dispersion relation for the excitations of the
inflaton\footnote{By inflaton we loosely understand the field generating
the primordial perturbations.}.  
The dispersion relation is
quadratic, $\omega^2\propto k^4$, in the case of the original ghost
inflation and linear, $\omega^2=\delta^2\cdot k^2$, with the small ``sound
speed'' $\delta\ll 1$ in the tilted version. This, combined with the
specific form of the interactions, leads to interesting predictions
for the amplitude and 
shape of non-Gaussianity, which have been constrained by the
Planck results \cite{Ade:2013ydc}.

The ghost inflation models are formulated as effective field theories (EFT)
for cosmological perturbations
valid below a certain cutoff. While this approach has the advantage of
being very general, it is unable to fully capture the evolution of the
inflaton background; in particular, it is problematic to incorporate in
it the end of inflation and reheating.
Besides, if one assumes that the ghost condensate
present at inflation persists till today, stability requires the
cutoff to be rather low
\cite{ArkaniHamed:2003uy,ArkaniHamed:2005gu}. This makes a
UV-completion 
of ghost inflation desirable.

In this paper we show
that ghost inflation can be embedded in the
framework of the khronometric or Einstein-aether gravity. The latter
being also an EFT, it has a cutoff as well. But this can
be as high as just a few orders of magnitude
below the Planck mass. In other words, khronometric
/ Einstein-aether can provide a partial UV-completion
(``UV-extension") for the ghost condensate
model almost up to the Planck scale. This allows to describe the
evolution of the background and perturbations within a self-contained
theory. 
Interestingly, the UV-extension
happens without restoration of the broken Lorentz symmetry, similar to
the case recently discussed in \cite{Endlich:2013vfa}. Further, in the
khronometric version of the setup there is a potential UV-completion
all the way above the Planck scale in the form of Ho\v rava gravity.

We study the observational signatures of the extended
model. For two reasons the analysis somewhat differs from the
discussion of the generic ghost inflation present in the
literature. First, the ability to keep the background evolution under
control allows to consider the situation when the VEV of the 
inflaton time derivative $\langle \dot\Theta\rangle$ --- the ``ghost
condensate'' ---  varies with time. This
produces a contribution into the tilt of the power spectrum in
addition to that coming from a potential for $\Theta$, which in some
cases can actually be dominant. Second, the ghost inflation is
characterized essentially by three energy scales: the scale
$\rho_{inf}^{1/4}$ of the inflationary energy density; the scale $M$
of the ghost condensate $M^2=\langle\dot\Theta\rangle$ which
determines the strength of the inflaton self-interaction
producing the
leading non-Gaussianity; and the scale $M'$ suppressing terms with
higher spatial derivatives in the quadratic effective action for
inflaton perturbations. 
In the previous treatments of
the ghost inflation the latter two scales have been commonly assumed to
be of the same order with the first scale being much higher,
$\rho_{inf}^{1/4}\gg M\sim M'$. We will see that in the extended
model where all the above scales are derived quantities, $M$ and $M'$ have
different parameter dependence, and the requirement that the EFT
description is valid for the inflationary background imposes the
hierarchy\footnote{At first sight, this hierarchy might seem
  surprising from the viewpoint of the EFT for the inflaton {\it
    perturbations}. However, it is straightforward to verify that it
  is
 stable under radiative corrections
  and thus perfectly natural.} 
$M\gg M'$. On the other hand, $\rho_{inf}^{1/4}$
can naturally be of the same order as $M$. As a result of this new
hierarchy, the amplitude of non-Gaussianity is somewhat suppressed
compared to the original prediction of the ghost inflation;
still, it remains large enough to be observationally interesting.

The paper is organized as follows. In Sec.~\ref{sec:fast} we describe
the model and identify the inflationary regime that reproduces 
ghost inflation. In Sec.~\ref{sec:lin} we study linear
cosmological perturbations emphasizing the similarities and
differences with the generic ghost
inflation treatment. Sec.~\ref{sec:bispectra} contains calculation of the
bispectrum. In Sec.~\ref{sec:kinetic} we apply our results to the
special case when the background dynamics is dominated by the kinetic
energy of the inflaton and derive observational constraints on the
model parameters in
this case. Sec.~\ref{sec:discussion} is devoted to conclusions. Some
details of the analysis are postponed to the
Appendices.

\section{Fast-roll inflation}
\label{sec:fast}

We start with the class of gravity theories containing in addition to
the metric $g_{\m\n}$ a dynamical time-like vector field $u_\m$ with
unit norm,\footnote{Our signature convention is $(+,-,-,-)$.}
\be 
u_{\mu}u^{\m}=1\,.
\ee
This field, called aether, 
exists in every point of space-time
and defines a preferred reference 
system, breaking the local LI of GR down to
the subgroup of spatial rotations around this vector. 
The dynamics of $u_\m$ is described by the most general covariant action
containing up to two derivatives,
\be
\label{aeact}
S_{[EH]}+S_{[u]}=-\frac{M_0^2}{2}\int \di^4x \sqrt{-g}
\big(R+K^{\m\n}_{\phantom{\m\n}\r\s}\nabla_\m u^\r\nabla_\nu u^\s\big)\;,
\ee
where 
\be
\label{Kmunulr}
K^{\mu\nu}_{\phantom{\m\n}\rho\s}=c_1 g^{\m\n}g_{\r\s}
+c_2\delta^\mu_\rho\delta^\nu_\sigma
+c_3\delta^\mu_\sigma\delta^\nu_\rho
+c_4 u^\mu u^\nu g_{\rho\s}\;,
\ee 
$c_1,\ldots,c_4$ are dimensionless parameters and $M_0$ is related to the Planck
mass,
\be
\label{PM}
M_P^2=M_0^2\bigg(1-\frac{c_1+c_4}{2}\bigg)\;. 
\ee
This class of theories has been introduced in \cite{Jacobson:2000xp}
(see also \cite{Jacobson:2008aj}) and received the name of
Einstein-aether models. 

It is possible to impose further restriction on $u_\m$ to make it
 hypersurface-orthogonal. This condition can be solved explicitly in
terms of a scalar field
\be
\label{Kh}
u_\m\equiv\frac{\nabla_\m \sigma}{\sqrt{\nabla^\n \sigma \nabla_\n \sigma}}\,,
\ee
where $\sigma(t,{\bf x})$ is assumed to have a non-vanishing time-like
gradient everywhere. Then out of the four terms with the derivatives of
$u_\m$ in (\ref{aeact}) only three are independent and the theory is
parameterized by the following combinations,
\be
\label{khpar}
\a\equiv c_1+c_4~,~~~\b\equiv c_1+c_3~,~~~\l\equiv c_2\;.
\ee
The geometrical meaning of $\s$ is that it labels the slices of a
preferred space-time foliation. From the physical viewpoint $\s$
sets a preferred time variable: hence the name ``khronon''. The theory
with the action (\ref{aeact}) and $u_\m$ expressed as in (\ref{Kh})
was introduced in \cite{Blas:2010hb} under the name ``khronometric''
model. Note that it is invariant  
under
reparameterizations of $\s$,
\be 
\label{repar}
\sigma \mapsto \tilde{\sigma}(\sigma)\,,
\ee
where $\tilde{\sigma}(\sigma)$ is an arbitrary monotonic function.
It has been shown to arise as the low-energy limit of Ho\v rava
gravity \cite{Horava:2009uw}. In other words, Ho\v rava gravity
can potentially provide a UV-completion to the model at 
trans-Planckian scales.

The generic Einstein-aether theory differs from the khronometric
version by the presence of vector modes, while the scalar and tensor
sectors in the two models are identical.\footnote{Another difference
  is the instantaneous interaction arising in the khronometric model
  \cite{Blas:2011ni}. However, this is irrelevant for the local
  physics studied in this paper.} We will see in the next section that
vector perturbations are not generated during inflation for the choice
of parameters we are interested in. Therefore, without loss of
generality, we will focus on the khronometric case and use the 
parameterization (\ref{khpar}).
We will assume throughout the paper
that the parameters $\a,\b,\l$ are of the same order and use $\a$ in
various estimates as a shorthand for the whole set $\a,\b,\l$.
Then the khronometric gravity is a valid EFT up to the 
scale\footnote{A nice study of this subject for the generic
  Einstein-aether theory is given in \cite{Withers:2009qg}.} 
\cite{Blas:2009qj,Blas:2010hb}, 
\be
\label{cutoff}
\Lambda_{\text{cutoff}}=M_0\sqrt{\alpha}\;,
\ee
which
is only a few orders below the Planck mass if $\a$ is not extremely
small. To avoid instabilities 
and negative energy the parameters must satisfy 
\cite{Blas:2009ck,Blas:2010hb}
\[
0<\a<2\,,\quad 0<\b+\l\,.
\]
Observations require $\a,\b,\l$ to be small. In particular, 
in the absence of fine-tuning between 
$\a,\b,\l$
the Solar
System constraints
\cite{Will:2005va} on the post-Newtonian parameters $\a_1^{PPN}$,
$\a_2^{PPN}$ describing LV  can be translated into the 
bounds, 
\be
\label{boundPPN}
|\a|,~|\b|,~|\l|\lesssim 10^{-7}\;.
\ee
Even stronger constraint on $\hat{\a}_2^{PPN}$ ---
the analog of $\a_2^{PPN}$ in strong gravitational field --- was
recently obtained from pulsar timing \cite{Shao:2013wga}.
However, we will disregard these constraints in this paper for two
reasons. First, a single fine-tuning $\a=2\b$ is sufficient to make
both PPN parameters vanish
 \cite{Blas:2010hb}.\footnote{This fine-tuning also greatly suppresses
   the strong-field PPN parameters as implied by the results of
  \cite{Yagi:2013ava}. We thank Diego Blas for the discussion of
  this point.} 
Then one is left with
much weaker bounds on $\a,\b,\l$ at the level of per cent from emission of the
gravitational waves \cite{Blas:2011zd,Yagi:2013qpa,Yagi:2013ava} and
late-time cosmology \cite{Audren:2013dwa}. Second, the values of
$\a,\b,\l$ during inflation can be different from the present epoch
(e.g., they can depend on the inflaton field) and may well exceed 
(\ref{boundPPN}). Thus, the only a priori assumption we are going to
make to simplify the calculations is $\a,\b,\l\ll 1$. This
will be validated at the end by the constraints on the
primordial spectrum following from the Planck data.

We now introduce the inflaton field $\Theta$. As it is well-known, to
sustain inflation the potential for $\T$ must be very flat which can be
achieved by imposing an approximate shift symmetry,
\be
\label{shift}
\T\mapsto \T+\text{const}\,.
\ee
Assume for a moment that this symmetry is exact. Then only derivatives
of $\T$ can appear
in the action. Allowing for the coupling
between $\T$ and the aether we obtain the most general action 
containing operators of dimension up to four,\footnote{As explained in
\cite{Blas:2011en}, the operators $u^\m\nabla_\m u^\n\,\d_\n\T$, 
$\nabla_\m u^\m\, u^\n\d_\n\T$ that naively have dimension 3, are in
fact of higher dimension when expressed in terms of the canonically
normalized fluctuations of the fields. Note that these oprators can be
forbidden by imposing the symmetry $u_\m\mapsto -u_\m$, $\T\mapsto
-\T$.} 
\be 
\label{inflAct}
S_{[\T]}=\int \di^4x
\sqrt{-g}\bigg[\frac{1}{2}g^{\m\n}\pd_\m\T\pd_\n\T
+\frac{\vk}{2}(u^\m\pd_\m\T)^2
-\mu^2\,u^\n\pd_\n\T -V\bigg]\,,
\ee
where $\vk$, $\m$ and $V$ are constants.
One will easily convince oneself that the cutoff of this action
combined with (\ref{aeact}) is still given by (\ref{cutoff}). Note
that it is technically natural for
 the mass parameter $\m$ to be smaller than
$\Lambda_\mathrm{cutoff}$ as it is protected from large quantum corrections
by the discrete symmetry $\T\mapsto -\T$ obeyed by the rest of the
action. 
The model with non-minimal coupling between the aether and the
inflaton similar to (\ref{inflAct}) was first introduced in
\cite{Donnelly:2010cr} and has been recently studied in
\cite{Solomon:2013iza}. In \cite{Blas:2011en,Audren:2013dwa} the model
with $V=0$ was suggested as a possible Lagrangian for dark energy
naturally protected from large quantum corrections. To avoid
confusion,
we point out that the model of this paper differs from that considered
in \cite{Creminelli:2012xb} where the khronon itself plays the role of
the inflaton.

Let us analyze the background cosmology in the constructed
model. Assuming the metric to be spatially flat we substitute the
homogeneous isotropic Ansatz,
\bseq
\label{backgrnd}
\begin{align}
&ds^2=N(t)^2dt^2-a(t)^2 {\bf dx}^2\;, \\
&u_0=N(t)~,~~~ u_i=0\;,\\
&\T=\T(t)\;,
\end{align}
\eseq  
into the action,
\be
S_{[EH]}+S_{[u]}+S_{[\T]}=\int
d^4 x\,a^3\;\bigg[-\frac{M^2_0(6+3\beta+9\lambda)}{2} 
\frac{\dot{a}^2}{Na^2}+\frac{{\dot\T}^2}{2Nc_\T^2}
-\mu^2\dot{\T} -V N\bigg]\,,
\ee
where we have introduced the notation 
\be
\label{cT}
c_\T^2=\frac{1}{1+\vk}\,.
\ee 
Varying this expression with
respect to $\T$ and $N$ we obtain,
\begin{align}
\label{eomPhi0}
&\frac{d}{dt}\bigg[a^3\,\bigg(\frac{\dot\T}{c_\T^2}-\m^2\bigg)\bigg]=0\;,\\
\label{Friedmann}
&H^2=\frac{1}{3M_0^2(1+\beta/2+3\lambda/2)}\left(
 \frac{\dot{\T}^2}{2c_\T^2}+V\right)\,,
\end{align}
where we fixed the gauge $N=1$ and introduced the Hubble
parameter $H\equiv \dot a/a$. From Eq.~(\ref{eomPhi0}) we see that the
time derivative of the inflaton develops a non zero VEV in the
stationary regime,
\be
\label{Tdot0}
\dot\T=\m^2c_\T^2\;.
\ee
This is precisely the type of the background relevant for the ghost
inflation. Note that the kinetic energy of $\T$ contributes into the
r.h.s. of the Friedmann equation (\ref{Friedmann}), which makes
accelerated expansion possible even in the absence of any potential,
$V=0$ \cite{Blas:2011en}. 

For inflation to end, the shift symmetry (\ref{shift}) must be
broken. We will incorporate this by promoting the parameters in
(\ref{inflAct}) to slowly varying functions\footnote{As noted above,
  the parameters of the aether / khronon Lagrangian (\ref{aeact}) can,
in principle, also change during or after inflation. Including their
dependence on the inflaton $\T$ is straightforward but renders the
formulas rather cumbersome. For the sake of clarity we will assume
that they stay constant during the period of inflation responsible for
the generation of the observed part of the perturbation spectrum.} of $\T$,
\be
\vk \mapsto \vk(\T)~,~~~\m\mapsto\m(\T)~,~~~
V\mapsto V(\T)\;.
\ee  
This does not affect the Friedmann equation (\ref{Friedmann}), while
the equation for $\T$ gets modified. Instead of (\ref{eomPhi0}) we
obtain,
\be
\label{eomPhi2}
\ddot\T+3H(\dot\T-\m^2c^2_\T)+c_\T^2V'-\frac{c'_\T}{c_\T}{\dot\T}^2=0\;,
\ee 
where prime stands for the derivative with respect to $\T$.
One can now identify two different inflationary regimes.

\textbf{1. Fast-roll inflation.}
It corresponds to the case when the contribution of the additional
terms related to the breaking of the shift symmetry is small, so that
the time derivative of the inflaton is still approximately given by
(\ref{Tdot0}). To find the conditions for the validity of this regime,
we write 
\be
\label{Tdot1}
\dot\T=\m^2c_\T^2(1+\delta^2)
\ee
with 
\be
\label{deltasmall}
\delta^2\ll 1
\ee
and substitute it into (\ref{eomPhi2}). Neglecting the terms with time
derivatives of $\delta^2$ (the validity of this approximation will be
discussed shortly) we obtain,
\be
\label{delta0}
\delta^2=-\frac{1}{3H}\bigg(2\m\m'c_\T^2+\m^2c_\T c_\T'
+\frac{V'}{\m^2}\bigg)\;.
\ee
The condition (\ref{deltasmall}) is satisfied provided that
\bseq
\label{c12}
\begin{align}
\label{c1}
&\frac{\m'}{\m}~,~\frac{c_\T'}{c_\T}\ll\frac{H}{\m^2 c_\T^2}\;,\\
\label{c2}
&V'\ll \m^2H\;.
\end{align}
\eseq
These inequalities have clear physical interpretation. The conditions
(\ref{c1}) can be written as $\frac{\dot\m}{\m},\frac{\dot
  c_\T}{c_\T}\ll H$, i.e. the variations of the parameters during the
Hubble time are small. The third inequality (\ref{c2}) states that the
contribution of the potential into the force acting on the inflaton is
small compared to the contribution which
arises from the coupling to the aether. 

Finally, in deriving
(\ref{delta0}) we have neglected the terms in
(\ref{eomPhi2}) with time-derivatives of $\delta^2$. It is straightforward
to see that this is justified as long as all quantities in
(\ref{delta0}) change slowly, as one expects during a quasi-de Sitter
stage. In addition to (\ref{c12}) this implies,
\be
\label{c3}
\frac{\m''}{\m'}~,~\frac{c_\T''}{c_\T'}~,~\frac{V''}{V'}\lesssim
\frac{H}{\m^2c_\T^2}\;. 
\ee 

An interesting limit of the fast-roll inflation arises when the
potential represents only a small correction to the kinetic energy or
is absent altogether,
\be 
\m^4\gg V\,.
\ee
Then the Hubble rate is directly related to the VEV of the inflaton
time derivative (or ghost condensate in the language of ghost
inflation),
\be 
\label{kindriven}
H^2\approx \frac{\m^4c_\T^2}{6M_0^2}\;.
\ee
We will call this limit 
\textbf{kinetically driven inflation} and will study it in detail in
Sec.~\ref{sec:kinetic}. 

\textbf{2. Slow-roll inflation.} This occurs when the inequality
(\ref{c2}) is replaced by the reverse,
\be
\label{cslow}
V'\gg \m^2 H\;.
\ee
In this case the background dynamics of the inflaton is essentially
the same as in the standard potentially-driven inflation. This regime
has been recently analyzed in detail in \cite{Solomon:2013iza} where
it was concluded that the aether-inflaton coupling either leads to
tachyonic instability of the cosmological perturbations or its effect
is unobservably small. In this paper we concentrate on the opposite
regime of the fast-roll where the behavior of
perturbations is quite different as we are going to see now.

\section{Linear perturbations}
\label{sec:lin}
We start the analysis of the perturbations by briefly discussing the
tensor modes. These are not modified by the introduction of the
inflaton $\T$ and behave exactly in the same way as in pure
Einstein-aether or khronometric theory. Thus we can use the known
result \cite{ArmendarizPicon:2010rs} that the only modification of the
primordial gravity wave power spectrum is the change of normalization
due to deviation of the velocity of tensor modes $c_t$ from
unity,\footnote{The factor 2 compared to Eq.~(29) 
of \cite{ArmendarizPicon:2010rs} comes from the sum over 
two polarizations of the
gravity waves.}
\be 
\label{tensors}
{\cal P}_h(k)=\frac{2}{\pi^2 c_t}\frac{H^2}{M_0^2}\bigg|_{c_t k=aH}\,,\quad
\,\text{where}\quad c_t^2=\frac{1}{1-\b}\, 
\ee  
and $k$ is the comoving wavenumber.
This is a small effect under our assumption $\b\ll 1$.

The analysis of other excitations is greatly simplified by the
observation that the dynamics of the aether (khronon) and inflaton
decouple from those of gravity in the regime of interest. To show this
we write
\be
\label{perts}
g_{\m\n}=\bar{g}_{\m\n}+h_{\m\n}~,~~~u_\m=\bar{u}_\m+v_\m~,~~~
\T=\bar\T+\pi\;,
\ee
where the background quantities $\bar g_{\m\n}$, $\bar u_\m$, $\bar\T$
are given by Eqs.~(\ref{backgrnd}) with $N(t)=1$. We want to compare
the terms mixing $h_{\m\n}$ with $v_\m$ and $\pi$ in the quadratic
Lagrangian to the kinetic terms for these fluctuations. Let us first
show that the mixing between $h_{\m\n}$ and $v_\m$ can be neglected
provided\footnote{Recall that $\a$ collectively denotes the order of
  magnitude of all coefficients $c_a$ in the aether Lagrangian.} 
$\a\ll 1$. Schematically one has the following estimates
\be
\label{hvmix}
{\cal L}^{(2)}_{[EH]}+{\cal L}_{[u]}^{(2)}
\sim \frac{M_0^2}{2}(\d h)^2+\frac{M_0^2\a}{2}[(\d v)^2
+\d v\d h+(\d h)^2]
\sim \frac{(\d\hat h)^2}{2}+\frac{(\d\hat v)^2}{2}
+\sqrt\frac{\a}{1+\a} \,\frac{\d\hat v\d\hat h}{2}\;, 
\ee 
where in the second expression we have introduced the canonically
normalized fields $\hat v_\m$, $\hat h_{\m\n}$. We see that, indeed,
the mixing term is suppressed by $\sqrt\a$.
The analysis of the inflaton--metric mixing requires more work as it
involves non-trivial cancellations between various terms. It is
performed in Appendix~\ref{app:A} where it is demonstrated that in
the leading approximation the mixing can be neglected. This means that
we can consistently set the metric to its unperturbed value and
consider only excitations of the aether/khronon and inflaton in the
external quasi-de Sitter background. In Appendix B we check explicitly
that the corrections to this approximation are small by 
computing the quadratic 
effective action for scalar inflationary perturbations in the
uniform inflaton gauge with fully dynamical metric.

The quadratic action for the perturbations reads,
\be
\label{S2FRW}
\begin{split}
S_{[u]}^{(2)}+S_{[\T]}^{(2)}=
\int \di^4x\,a^3\,&\bigg[M_0^2(c_1+c_4)\frac{\dot v_i^2}{2a^2}
-M_0^2c_1\frac{\d_i v_k\d_i v_k}{2a^4}
-M_0^2(c_2+c_3)\frac{(\d_i v_i)^2}{2a^4}\\
&+\frac{(1+\vk)\dot{\pi}^2}{2}
-\frac{(\d_i\pi)^2}{2a^2}
+(\m^2-\vk\dot{\bar\T})\frac{v_i\d_i\pi}{a^2}
-(\m^2-\vk\dot{\bar\T})\dot{\bar\T}\frac{v_i^2}{2a^2}
\bigg]\;.
\end{split}
\ee
In deriving this expression we have neglected several types of
contributions. First, we neglected the terms coming from the Taylor
expansion of the functions $\vk(\T)$, $\m(\T)$ and $V(\T)$. These
terms give an effective mass for the field $\pi$ which, due to the
conditions (\ref{c12}), (\ref{c3}) is much smaller than the Hubble
rate. Thus they are irrelevant for the calculation of the primordial
power spectrum determined by the dynamics of the perturbations inside
the Hubble radius and at horizon crossing. Second, we discarded
terms of the form 
\[
O(\a) M_0^2H^2 v_i^2/a^2
\] 
appearing in the aether Lagrangian. These are smaller than the last
term in (\ref{S2FRW}) provided
\be
\label{condH}
H^2\ll\frac{\m^4}{M_0^2\a}\;.
\ee
For the kinetically driven inflation this condition is automatically
fulfilled. We will assume it to be valid also in the presence of the
potential, as only in this case we obtain the interesting
phenomenology of the ghost inflation. 

Following the standard procedure one decomposes $v_i$ into the
transverse and longitudinal parts,
\be
\label{vdecomp}
v_i=v_i^{(t)}+\d_i\chi~,~~~\d_i v_i^{(t)}=0\;.
\ee 
Note that the transverse part is present only in the case of the
general aether, while it identically vanishes in the khronometric
case. In the former case the equation for it follows from
(\ref{S2FRW}), 
\be
\label{eqvect}
{\ddot v}_i^{(t)}+H {\dot v}_i^{(t)}
-\frac{c_1}{c_1+c_4}\frac{\D v_i^{(t)}}{a^2}
+\frac{\m^4 c_\T^4}{M_0^2(c_1+c_4)}v_i^{(t)}=0\;,
\ee
where $\D\equiv\d_i\d_i$. We observe that the last term gives a mass
to the transverse vector modes which, according to (\ref{condH}), is
much larger than the Hubble rate (we assume throughout the paper that
$c_\T$ is of order 1). This implies that these modes are not generated
during inflation which justifies our previous assertion. Thus for the
purposes of this paper the general aether is equivalent to its
khronometric reduction. 

Henceforth we concentrate on the scalar sector of perturbations. Note
that in the khronometric model $\chi$ coincides with the perturbation
of the khronon field which we write as,\footnote{The background value
  of the khronon can always be brought to the form $\bar\sigma=t$
  using the reparameterization (\ref{repar}).}
\be
\label{khrpert}
\sigma=t+\chi\;.
\ee 
The equations for the coupled system of the fields $\chi$ and $\pi$ are,
\bseq
\label{coupled}
\begin{align}
\label{chipi}
&\ddot{\chi}+H\dot{\chi}-c_\chi^2\frac{\Delta \chi}{a^2}
+\frac{\m^2-\vk\dot{\bar\T}}{M_0^2\a}(\dot{\bar\T}\chi-\pi)=0\;,\\
\label{pichi}
&\ddot{\pi}+3H\dot{\pi}-c_\T^2\frac{\Delta \pi}{a^2} 
+(\m^2-\vk\dot{\bar\T})c_\T^2\frac{\D\chi}{a^2}=0\;,
\end{align}
\eseq
where we have introduced the notation
\be
\label{cchi}
c_\chi^2=\frac{\b+\l}{\a}\;.
\ee
Let us study the subhorizon behavior of the modes. Neglecting the
terms containing $H$ and performing the Fourier decomposition, 
$
\chi, \pi \propto e^{-i\omega t+i{\bf kx}}\,,
$
we obtain the dispersion relations,
\be 
\label{disp1}
\omega^2_\pm= \frac{c_\chi^2+c_\T^2}{2}\bigg[
\frac{k^2}{a^2}+k_c^2 
\pm  \sqrt{\bigg(\frac{k^2}{a^2}+k_c^2 \bigg)^2
-\frac{4c_\T^2 c_\chi^2}{(c_\chi^2+c_\T^2)^2}\frac{k^4}{a^4}
-\frac{4\delta^2 k_c^2}{c_\chi^2+c_\T^2}\frac{k^2}{a^2}\bigg]}\;,
\ee 
where
\be
\label{kc}
k_c^2\equiv\frac{\m^4c_\T^4}{M_0^2(c_\chi^2+c_\T^2)\a}\;.
\ee
Note that (\ref{condH}) implies $k_c\gg H$. At large momenta, $k/a\gg
k_c$, the fields $\chi$ and $\pi$ decouple and the dispersion
relations become linear,
\be
\omega^2_{+}=c_\chi^2 (k/a)^2 \,,\quad 
\omega^2_{-}=c_\T^2 (k/a)^2\,,
\ee
where the velocities $c_\chi$, $c_\T$ in general differ from unity
as the consequence of the Lorentz violation. At small momenta,
$k/a\ll k_c$, on the contrary, the mixing is large and the dispersion
relations take the form, 
\begin{align}
\label{mness}
\omega^2_{+}&=(c_\chi^2+c_\T^2)k_c^2+(c_\chi^2+c_\T^2-2\delta^2)(k/a)^2\,,\\
\label{mless}
\omega^2_{-}&=\delta^2 \,(k/a)^2+\frac{c_\chi^2c_\T^2}{c_\chi^2+c_\T^2}
\frac{(k/a)^4}{k_c^2}\,,
\end{align}
where $\delta$ is given by Eq.~(\ref{delta0}). 
From (\ref{mness}) we see that one family of perturbations
acquires an energy gap much larger than the Hubble rate. Thus it is
irrelevant for the inflationary physics and can be integrated
out. Effectively we are dealing with a single field
inflation\footnote{This implies absence of isocurvature perturbations.}
described by the second mode which remains gapless, see
(\ref{mless}). Remarkably, the linear term in its dispersion relation
is strongly suppressed by the small parameter $\delta^2$; in the limit
of the exact shift symmetry of the inflaton the linear contribution
disappears altogether and one is left with the quadratic dispersion
relation $\omega^2\propto k^4$. This exactly coincides with the
situation in the tilted ghost inflation
\cite{ArkaniHamed:2003uz,Senatore:2004rj}. Note, however, that our
setup violates an assumption often present in the discussion of the ghost
inflation that the scale suppressing the $k^4$ term in the dispersion
relation is of the same order as the scale of the ghost condensate. In
our case we have from (\ref{Tdot1}), (\ref{kc}),
\be
k_c\sim \frac{\dot{\bar\T}}{M_0\sqrt{\a}}\;.
\ee  
The validity of the EFT description for the background requires $\sqrt
{\dot{\bar\T}}$ to be smaller than the cutoff of the theory
  (\ref{cutoff}), which implies the hierarchy
\be
\label{hier}
k_c\ll \sqrt{\dot{\bar\T}}\;.
\ee
We emphasize that this hierarchy naturally appears in the framework
which captures both the background evolution and perturbations within
a single consistent EFT. 

To find the low-energy effective action for the gapless perturbations
one notes that at $k/a\ll k_c$ equation (\ref{chipi}) is dominated by
the last term which tightly couples $\chi$ to $\pi$,
\be
\label{tracer}
\chi=\pi/\dot{\bar\T}\;.
\ee
Substituting this back into (\ref{S2FRW}) yields the quadratic action,
\be
\label{gheff}
S_{[\pi]}^{(2)}=\int \di^4x\, \frac{a^3}{2c_\T^2}
\bigg[{\dot{\pi}}^2-\delta^2\frac{(\d_i\pi)^2}{a^2}
-\frac{c_{\chi}^2c_\T^2}{c_\chi^2+c_\T^2}\frac{(\D\pi)^2}{k_c^2a^4}
\bigg]\;,
\ee
where we have neglected the terms containing spatial derivatives of
$\dot\pi$ as they are irrelevant at low momenta. Clearly, this action
reproduces the dispersion relation (\ref{mless}). The subsequent
analysis proceeds differently depending on whether the dispersion
relation is dominated by the first or second term in (\ref{mless})
when the mode freezes out, which occurs when the frequency of the mode
drops down to the Hubble rate, $\omega\sim H$.

\paragraph{Quadratic dispersion relation}~\\
By inspection of Eq.~(\ref{mless}) one finds that the first term on
the r.h.s. is never important if
\be
\label{quadrcond}
\delta^2\ll H/k_c\;.
\ee
Then the mode equation following from (\ref{gheff}) has the form,
\be 
\label{ghosteqm}
\ddot{\pi}_{\bf k}+3H\dot{\pi}_{\bf k} +\frac{c_\chi^2c_\T^2}{c_\chi^2+c_\T^2} 
\frac{k^4}{k_c^2a^4}\pi_{\bf k}=0\;.
\ee
One quantizes the $\pi$-field,
\be
\label{ghFourie}
\pi({\bf x},t)=\int \frac{d^3k}{(2\pi)^{3}}\big(\pi_{\bf k}(t)
a_{\bf k}+\pi_{\bf k}^*(t)
a^{+}_{-{\bf k}}\big)e^{i{\bf k x}}\,,
\ee
with the creation-annihilation operators obeying the standard
commutation relations 
\[
[a_{\bf k},a^{+}_{\bf q}]=(2\pi)^3\delta^{3}({\bf k}-{\bf q})\,.
\]
The mode functions $\pi_{\bf k}(t)$ are the positive-frequency
solutions of Eq.~(\ref{ghosteqm}),
\be 
\label{wfgen}
\pi_{\bf k}=\sqrt{\frac{\pi}{8}}c_\T H |\eta|^{3/2} 
H^{(1)}_{3/4}\bigg(\frac{c_{\chi}c_\T
  H}{2(c^2_\chi+c_\T^2)^{1/2}k_c}k^2\eta^2
\bigg)\,,
\ee
where $\eta=\int dt/a$ is the conformal time and $H_{3/4}^{(1)}$ is
the Hankel function. The normalization of (\ref{wfgen}) is fixed by
imposing the canonical commutation relations on $\pi({\bf x},t)$ and
its conjugate momentum following from (\ref{gheff}). The power
spectrum of $\pi$ is given by the formula
\be
\label{pwpi}
{\cal P}_{\pi}=\frac{k^3}{2\pi^2}\lim_{\eta\to -0}|\pi_{\bf k}(\eta)|^2. 
\ee

We are interested in the power spectrum of the gauge invariant
perturbation $\zeta$ which stays constant outside the horizon 
\cite{Bardeen:1983qw,Salopek:1990jq}.
This is defined as the fluctuation of the logarithm of the scale
factor on the surfaces of constant inflaton field.
In our 
approximation of the decoupling between the inflaton and the metric
perturbations and neglecting
small deviations of the background
space-time from de Sitter $\zeta$ is related to $\pi$ by a simple
formula\footnote{For the analysis in the next section it is important
  to stress that within the above approximation this relation holds at
  non-linear order \cite{ArkaniHamed:2003uz}.} 
(cf.~\cite{Cheung:2007st}),
\be
\label{zeta}
\zeta=
-\frac{H}{\dot{\bar\T}}\pi\;.
\ee 
Combining everything together and using the Taylor expansion for the
Hankel functions we obtain,
\be 
\label{Pow2}
{\cal P}_{\zeta}(k)=\frac{1}{\pi
  (\Gamma(1/4))^2}\cdot\frac{1}{c_\chi^{3/2}c_\T^{1/2}}
\cdot \frac{H^{5/2}}{\mu M_0^{3/2}\alpha^{3/4}}
\bigg|_{a^2H=\frac{c_\chi M_0\sqrt{\a}}{c_\T\mu^2}k^2}\;,
\ee
where we have emphasized explicitly that all the quantities entering
into this expression must be evaluated at the moment when the mode 
freezes out. The slow evolution of these quantities gives a tilt of the
power spectrum,
\be 
\label{tilt2}
n_s-1\equiv \frac{d\log{\cal P}_\zeta}{d\log k}
=-\frac{\m^2 c_\T^2}{2H}\bigg(\frac{c_\T'}{c_\T}+\frac{2\m'}{\m}\bigg)
-\frac{5\m^4 c_\T^2}{4M_0^2 H^2}\delta^2\;.
\ee
Finally, the tensor-to-scalar ratio is obtained using (\ref{tensors}),
\be
\label{scalar-tensor2}
r\equiv\frac{{\cal P}_t}{{\cal P}_\zeta}
=\frac{2(\Gamma(1/4))^2}{\pi}\cdot
\frac{c_\chi^{3/2}c_\T^{1/2}\m}{M_0^{1/2}H^{1/2}}\cdot\alpha^{3/4}\,,
\ee 
where we have neglected the difference of the graviton speed from 1.
In Sec.~\ref{sec:kinetic} we will use these expressions to make
numerical estimates in the special case of the kinetically-driven inflation. 

\paragraph{Linear dispersion relation}~\\
If 
\be
\label{linearcond}
\delta^2\gg H/k_c
\ee
the dispersion relation of the modes (\ref{mless}) is dominated by the
linear term at the moment of freeze-out. This means that we can
neglect the last term in the action (\ref{gheff}) when deriving the
mode equation,
\be
\ddot\pi+3H\dot\pi+\frac{k^2\delta^2}{a^2}\pi=0\;.
\ee
Its normalized positive frequency solution is 
\be
\label{wftilted}
\pi_{\bf k}=\frac{H c_\T}{\sqrt{2 k \delta}}|\eta|
\bigg(1-\frac{i}{k\eta\delta}\bigg)e^{-ik \eta\delta}\,.
\ee
This gives the power spectrum,
\be
\label{Powlin}
{\cal P}_\zeta(k)=
\frac{1}{4\pi^2}\cdot\frac{H^4}{\mu^4c_\T^2\delta^3}
\bigg|_{aH=k\delta}\,,
\ee
the tilt,
\be 
\label{tiltlin}
n_s-1
=-\frac{2\m^2 c_\T^2}{H}\bigg(\frac{c_\T'}{c_\T}+\frac{2\m'}{\m}\bigg)
-\frac{2\m^4 c_\T^2}{M_0^2 H^2}\,\delta^2-\frac{3\dot\delta}{H\delta}\;
\ee
and the tensor-to-scalar ratio,
\be
\label{scalar-tensorlin}
r=\frac{8\m^4 c_\T^2}{M_0^2 H^2}\cdot\delta^3\,.
\ee

\section{Bispectrum}
\label{sec:bispectra}

The interesting feature of ghost inflation is enhanced
non-Gaussianity of a special shape. In this section we are going to
show that this feature is reproduced in our model by computing the
three-point function of the curvature perturbations.

As a first step we must identify the leading interaction terms. To
this end we expand the aether and inflaton Lagrangians to the cubic
order in the perturbations. This task is simplified by several
observations. First, due to the decoupling of gravity, the metric
perturbations are not excited and one can set them to zero. Second,
the vector modes of the aether are not excited either as they are
massive during inflation. Thus the aether is reduced to its
longitudinal --- khronon --- part. 
Third, for an estimate we drop terms with time derivatives of the
scale factor arising in the expansion of the aether action; this is
a valid approximation\footnote{For the inflaton action we do not need
  this simplification as no derivatives of the scale factor appear in
  it. Thus Eq.~(\ref{LTh3}) is valid also for superhorizon modes.} 
for subhorizon modes with $\omega\gtrsim H$. 
 Finally, we neglect
contributions coming from the expansion of the functions $\vk(\T)$,
$\m(\T)$, $V(\T)$ as, by assumption, the deviation of the space-time
from de Sitter is small and these contributions are suppressed. In
this way one obtains,\footnote{To simplify
  (\ref{Lu3}) we have integrated by parts in the action. Note that the
term with higher time derivatives in (\ref{Lu3}) can be removed by the
field redefinition \cite{Blas:2010hb} 
$\chi=\bar\chi+\bar\chi\dot{\bar\chi}$
and thus does not lead to any new degrees of freedom.}
\begin{align}
\label{Lu3}
&{\cal
  L}_{[u]}^{(3)}=M_0^2\a\bigg(\frac{\dot\chi\d_i\ddot\chi\d_i\chi}{a^2}
-\frac{\d_i\dot\chi\d_j\chi\d_i\d_j\chi}{a^4}\bigg)
+\frac{M_0^2(\b+\l)}{a^4}
\big(2\d_i\dot\chi\d_i\chi\D\chi+\dot\chi(\D\chi)^2\big)\;,\\
\label{LTh3}
&{\cal L}_{[\T]}^{(3)}=(\m^2-\vk\dot{\bar\T})\dot{\bar\T}
\frac{\dot\chi(\d_i\chi)^2}{a^2}
-\bigg(\frac{\m^2}{2}-\vk\dot{\bar\T}\bigg)
\frac{(\d_i\chi)^2\dot\pi}{a^2}
-(\m^2-\vk\dot{\bar\T})
\frac{\dot\chi\d_i\chi\d_i\pi}{a^2}
-\vk\frac{\d_i\chi\dot\pi\d_i\pi}{a^2}\;.
\end{align}
We want to compare various terms in (\ref{Lu3}), (\ref{LTh3}) at the
time when the perturbations freeze out. The frequency and
momentum of the mode then satisfy 
\be
\label{frek}
H\sim\omega\ll k/a\ll k_c\;,
\ee
where the last two inequalities follow from the condition
(\ref{condH}) and the dispersion relation (\ref{mless}). This implies
that the
spatial gradients are enhanced relative to the time derivatives. Using
also the relation (\ref{tracer}) between $\chi$ and $\pi$ valid at low
frequency we obtain the estimates,
\be
\label{L3estims}
{\cal L}_{[u]}^{(3)}\sim\frac{M_0^2\a}{\m^6}\omega(k/a)^4\pi^3~,~~~~~
{\cal L}_{[\T]}^{(3)}\sim\frac{1}{\m^2}\omega(k/a)^2\pi^3\;.
\ee 
From (\ref{frek}) one immediately concludes 
${\cal L}_{[u]}^{(3)}\ll {\cal L}_{[\T]}^{(3)}$. In other words, the
cubic interaction is dominated by the contribution from the inflaton
Lagrangian. Expressing $\chi$ through $\pi$ and substituting the
background value $\dot{\bar\T}$ one ends up with 
\be
\label{Hint}
S_{[\pi]}^{(3)}=-\int d^4x\; \frac{a}{2\m^2c_\T^4}\,
\dot{\pi}(\d_i \pi)^2\;.
\ee
This has precisely the same form as in the ghost inflation model 
\cite{ArkaniHamed:2003uz}.

The rest of the analysis of the bispectrum proceeds along the lines of
\cite{ArkaniHamed:2003uz,Senatore:2004rj}.
To get an order-of-magnitude estimate for the size of non-linear
effects consider the ratio between the cubic and quadratic Lagrangians
at the moment of freeze-out,
\be
\label{3to2}
\frac{{\cal L}_{[\pi]}^{(3)}}{{\cal L}_{[\pi]}^{(2)}}
\sim \frac{\omega(k/a)^2\pi^3/\m^2}{\omega^2\pi^2}
\sim \frac{(k/a)^2}{\m^2 H}\pi\;.
\ee
The amplitude of the bispectrum is characterized by the quantity
$f_{NL}$ which is related to (\ref{3to2}) by \cite{Cheung:2007st}, 
\be
\label{NGestim}
|f_{NL}|\sim\frac{1}{\zeta}\frac{{\cal L}_{[\pi]}^{(3)}}{{\cal
    L}_{[\pi]}^{(2)}} \sim \frac{(k/a)^2}{H^2}\bigg|_\text{freeze-out}\;,
\ee
where we have used the relation
(\ref{zeta}) between $\zeta$ and $\pi$.
Depending on whether the dispersion relation of the perturbations is
dominated by the quadratic or linear term, we obtain,
\be
\label{NGestim1}
|f_{NL}|\sim\begin{cases}
k_c/H&\text{if}~~ \delta^2\ll H/k_c\\
1/\delta^2 &\text{if}~~ \delta^2\gg H/k_c
\end{cases}
\ee
In both cases the non-Gaussianity is parametrically 
enhanced, $|f_{NL}|\gg 1$. We point out, however, that it is smaller
than in the original ghost inflation model. To see this let us rewrite
(\ref{NGestim1}) using the expressions for the power spectrum
(\ref{Pow2}), (\ref{Powlin}). Dropping the numerical coefficients we
obtain, 
\be
\label{NGestim2}
|f_{NL}|\sim\begin{cases}
\zeta^{-4/5}\,(k_c/\m)^{8/5}&\text{if}~~ \delta^2\ll \zeta^{4/5}(\m/k_c)^{8/5}\\
1/\delta^2 &\text{if}~~ \delta^2\gg \zeta^{4/5}(\m/k_c)^{8/5}
\end{cases}
\ee
The predictions of
the original ghost inflation are recovered in the limit $k_c\sim
\mu$. The hierarchy of scales (\ref{hier}) clearly
introduces a suppression.

The precise expression for the bispectrum is obtained by evaluating
the three-point function of the inflaton fluctuation $\pi$ at the
moment of freeze-out $t_f$ and then switching to $\zeta$ using
(\ref{zeta}). Using the in-in perturbation theory
\cite{Maldacena:2002vr} with the interaction Lagrangian (\ref{Hint})
one obtains for the correlator of three Fourier harmonics,
\be
\label{Bs1}
\begin{split} 
\langle \pi({\bf k}_1, t_f)\pi({\bf k}_2, t_f)\pi({\bf k}_3, t_f)\rangle 
= i\int_{-\infty}^{t_f}&dt d^3x\,\frac{-a(t)}{2\m^2c_\T^4}\\
&\times\langle\, \big[\pi({\bf k}_1, t_f)\pi({\bf k}_2, t_f)\pi({\bf k}_3, t_f),\; 
\dot\pi({\bf x},t)\big(\d_i\pi({\bf x},t)\big)^2\big]\,\rangle_0 \,,
\end{split}
\ee
where on the r.h.s. the expectation value is evaluated in the
unperturbed vacuum. We are interested in the function $B_\zeta$ defined
as,
\be
\label{Bdef}
\langle \zeta({\bf k}_1) \zeta({\bf k}_2) \zeta({\bf k}_3)\rangle
=(2\pi)^3 \delta({\bf k}_1+{\bf k}_2+{\bf k}_3)\,
B_\zeta(k_1,k_2,k_3)\;, 
\ee
where $\delta$-function appears due to the translation invariance while
rotation symmetry implies that $B_\zeta$ depends only on the absolute
values of the momenta. Substituting the decomposition (\ref{ghFourie})
into (\ref{Bs1}) one obtains after a bit of algebra,
\be
\label{Bs2}
\begin{split}
B_\zeta(k_1,k_2,k_3) =& i\frac{H^2}{\m^8c_\T^{10}}
({\bf k}_2\cdot {\bf k}_3)
\lim_{\eta_f\to 0^-}\pi_{{\bf k}_1}(\eta_f)\pi_{{\bf k}_2}(\eta_f)
\pi_{{\bf k}_3}(\eta_f)\\
&\times\Re \int_{-\infty}^{\eta_f}\frac{d\eta}{\eta} 
\frac{d\pi_{{\bf k}_1}^*(\eta)}{d\eta}
\pi_{{\bf k}_2}^*(\eta) \pi_{{\bf k}_3}^*(\eta)
+ \text{perm.}\,,
\end{split}
\ee
where we have switched to integration over conformal time and
``perm.'' stands for the terms differing by permutations of the
momenta. The final answer depends on the precise form of the 
modes at the freeze-out.

\begin{figure}[tb]
\begin{center}
\includegraphics[width=0.45\textwidth]{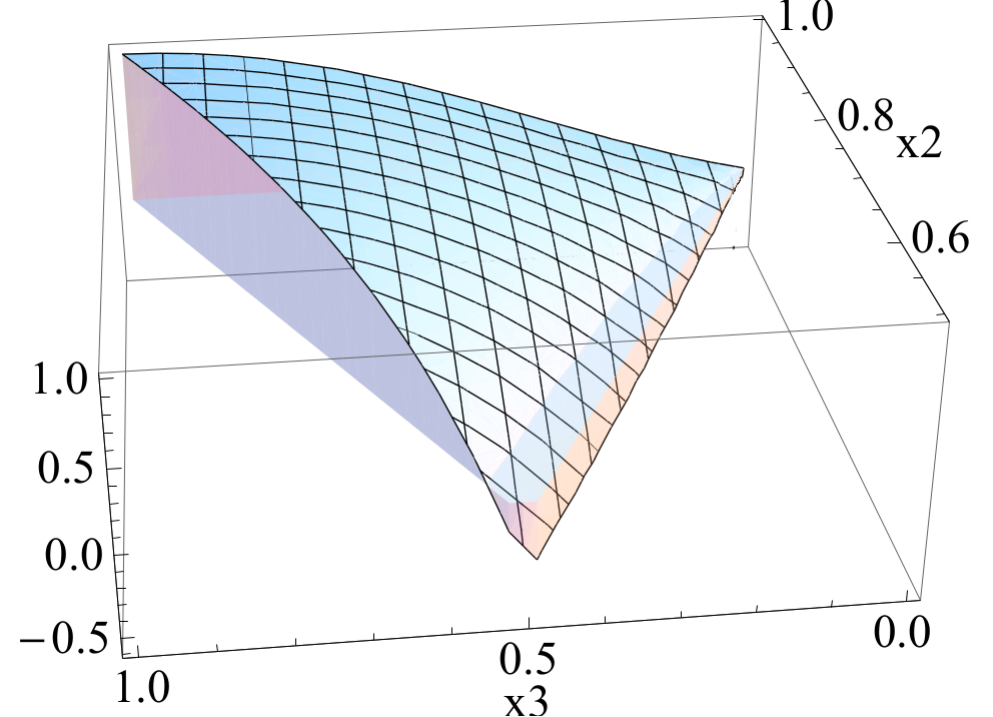}\qquad\quad
\includegraphics[width=0.45\textwidth]{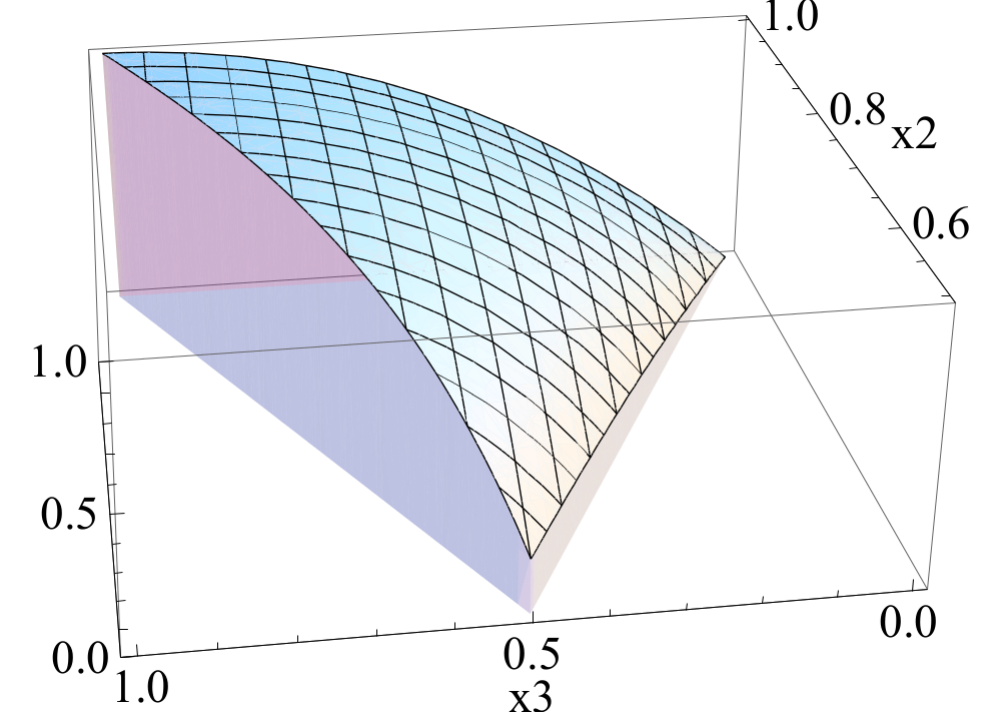}
\caption{
The shapes of the bispectrum $B_\zeta(1,x_2,x_3)x_2^2x_3^2$ for the
cases of quadratic 
(left panel), 
and linear dispersion relations (right panel). 
The functions have been normalized to one at $x_2=x_3=1$ and are
plotted only for inequivalent configurations of momenta 
$1-x_2\leq x_3\leq x_2$.
\label{Fig:1}
}
\end{center}
\end{figure}

The case of {\bf quadratic dispersion relation} yields,
\be
\label{BS22}
B_\zeta(k_1,k_2,k_3)=-\frac{H^4}{M_0^4c_\chi^4\a^2}\cdot
\frac{\pi^3}{16(\Gamma(1/4))^3}\,
\frac{k_1(k_1^2-k_2^2-k_3^2)}{(k_1k_2k_3)^3}\,
I\bigg(\frac{k_2}{k_1},\frac{k_3}{k_1}\bigg)+\mathrm{perm.}\;,
\ee
where
\be
\label{II}
I(x_2,x_3)=\Re\int_{-(1+i\epsilon)\infty}^0
\frac{dy}{y}\;\frac{dF(y)}{dy} F(x_2y) F(x_3y)
\ee
and
\be
F(y)=(-y)^{3/2}H_{3/4}^{(1)}(y^2/2)\;.
\ee
Note that in (\ref{II}) we have rotated the contour of integration
below the real axis to make the integral convergent. The shape of the
bispectrum obtained by numerically computing (\ref{II}) is shown in
Fig.~\ref{Fig:1} (left panel) as the function of two ratios, 
$x_{2,3}=k_{2,3}/k_1$ \cite{Babich:2004gb}. Clearly, it is maximal for
equilateral configurations $k_1\sim k_2\sim k_3$. 

To quantify its amplitude one evaluates the bispectrum at
equal momenta and introduces the parameter\footnote{This definition 
  of $f_{NL}$ agrees with the conventions used by the Planck
  collaboration \cite{Ade:2013ydc} upon substitution $\Phi=3\zeta/5$,
$P_\Phi=2\pi^2{\cal P}_\Phi/k^3$.} $f_{NL}$,
\be
\label{fNLdef}
B_\zeta(k,k,k)=\frac{3}{5}\cdot 6\cdot
\bigg(\frac{2\pi^2}{k^3}{\cal P}_\zeta(k)\bigg)^2\;f_{NL}\;.
\ee
Comparison with (\ref{BS22}), (\ref{Pow2}) yields,
\be
\label{fNL2}
f_{NL}=\frac{5\pi\Gamma(1/4)}{192} I(1,1)
\frac{\mu^2c_\T}{HM_0c_\chi\sqrt{\a}}
\approx
-0.13 \frac{\mu^2c_\T}{HM_0c_\chi\sqrt{\a}}\;.
\ee
Note that $f_{NL}$ is predicted to be negative. By the absolute value
(\ref{fNL2}) agrees with the estimate (\ref{NGestim1}). Notice though
a numerical factor giving suppression by about and order of magnitude.

In the case of {\bf linear dispersion relation} we obtain,
\be
\label{BS11}
B_\zeta(k_1,k_2,k_3)=\frac{H^8}{\m^8c_\T^4\delta^8}
\cdot\frac{1}{16}\,\frac{k_1^2(k_1^2-k_2^2-k_3^2)}{(k_1k_2k_3)^3}\,
\frac{k_1^2+2k_2^2+2k_3^2+3k_1k_2+3k_1k_3+6k_2k_3}{(k_1+k_2+k_3)^3}
+\mathrm{perm.}
\ee
The momentum dependence can be cast into the form,
\be
B_\zeta(k_1,\!k_2,\!k_3)\!\propto\!
\frac{\sum_a k_a^6+\sum_{a\neq b}(3k_a^5k_b\!-\!k_a^4k_b^2\!-\!3k_a^3k_b^3)
+\sum_{a\neq b\neq c}(3k_a^4k_bk_c\!-\!9k_a^3k_b^2k_c\!-\!2k_a^2k_b^2k_c^2)}
{(k_1k_2k_3)^3(k_1+k_2+k_3)^3}\;,
\ee
which is one of the templates used by the Planck collaboration in the
data analysis, see \cite{Ade:2013ydc} where it is called ``EFT1''. The
shape (\ref{BS11}) is plotted on the right panel of
Fig.~\ref{Fig:1}. It is similar to the previous case, but there are
visible differences. From Eqs.~(\ref{BS11}), (\ref{Powlin}) one
infers,
\be
\label{fNL1}
f_{NL}=-\frac{85}{324}\cdot\frac{1}{\delta^2}
\approx-\frac{0.26}{\delta^2}\;. 
\ee
Again, we see that $f_{NL}$ is negative and agrees with the estimate
(\ref{NGestim1}). Remarkably, in this case it depends on the single
parameter $\delta^2$.

\section{Kinetically driven inflation}
\label{sec:kinetic}

If the potential for the $\T$-field is absent, $V=0$, the Hubble rate
is given by the expression (\ref{kindriven}) which simplifies the
formulas and makes the model highly predictive. The condition
(\ref{quadrcond}) for the dominance of the quadratic piece in the
dispersion relation translates into
\be
\label{quadrcond1}
\delta^2\ll\sqrt\a
\ee 
and we obtain for the main observables,\\
\bseq
\label{qvalues}
\begin{minipage}{4cm}
\[
\begin{minipage}{2cm}
{\it quadratic}
{\it dispersion}
\end{minipage}
\begin{cases}
~\\
~\\
~\\
~\\
\\
\end{cases}
\]
\end{minipage}
\begin{minipage}{12.3cm}
\begin{align}
\label{qvalues1}
{\cal P}_\zeta&= 2.6\cdot 10^{-3}\;\frac{c_\T^2}{c_\chi^{3/2}}\,
\frac{\m^4}{M_0^4\a^{3/4}}\;,\\
\label{qvalues2}
n_s-1&=-6\,\delta^2\;,\\
\label{qvalues3}
r&= 13 c_\chi^{3/2} \alpha^{3/4}\;, \\
\label{qvalues4}
f_{NL}&=-\frac{0.32}{c_\chi\sqrt{\a}}\;.
\end{align}
\end{minipage}\\
\eseq
\\In the opposite case
\be
\label{lincond1}
\delta^2\gg\sqrt\a
\ee 
 corresponding to the linear dispersion relation 
we have\\
\bseq
\label{linvalues}
\begin{minipage}{4cm}
\[
\begin{minipage}{2cm}
{\it linear}\\
{\it dispersion}
\end{minipage}
\begin{cases}
~\\
~\\
~\\
~\\
\\
\end{cases}
\]
\end{minipage}
\begin{minipage}{12.3cm}
\begin{align}
\label{linvalues1}
{\cal P}_\zeta&= 0.7\cdot 10^{-3}\;
\frac{\m^4c_\T^2}{M_0^4}\,\frac{1}{\delta^3}\;,\\
\label{linvalues2}
n_s-1&=-6\,\delta^2-\frac{3\dot\delta}{H\delta}\;,\\
\label{linvalues3}
r&= 48\,\delta^3\;,\\
\label{linvalues4}
f_{NL}&=-\frac{0.26}{\delta^2}\;.
\end{align}
\end{minipage}
\eseq
These expressions should be compared with the best-fit values for the
cosmological parameters reported by the Planck collaboration 
\cite{Ade:2013zuv,Ade:2013uln},
\bseq
\label{cosmresults}
\begin{align}
\label{cosmresults1}
\ln(10^{10}{\cal P}_\zeta)\Big|_{k=0.05\mathrm{Mpc}^{-1}}&
= 3.089^{+0.024}_{-0.027}& (68\%~\mathrm{CL})\;,\\
\label{cosmresults2}
n_s&=0.9603\pm 0.0073& (68\%~\mathrm{CL})\;,\\
\label{cosmresults3}
r&< 0.11& (95\%~\mathrm{CL})~\; 
\end{align}
\eseq
and the constraints on the two relevant forms of non-Gaussianity ---
the ``ghost inflation'' shape and the ``EFT1'' shape in the
notations of \cite{Ade:2013ydc},
\bseq
\label{fNLresults}
\begin{align}
\label{fNLresults1}
f_{NL}^{\rm ghost}&
= -23\pm 88& (68\%~\mathrm{CL})\;,\\
\label{fNLresults2}
f_{NL}^{\rm EFT1}&
= 8\pm 73& (68\%~\mathrm{CL})\;.
\end{align}
\eseq

We first analyze the case of {\bf quadratic dispersion relation}. From
(\ref{qvalues2}), (\ref{cosmresults2}) one finds,
\be
\label{qdelta}
\delta^2\approx 6.6\cdot 10^{-3}\;.
\ee
Then Eq.~(\ref{quadrcond1}) gives us the condition for the validity of
the regime at hand,
\be
\label{qalphalow}
\a\gg 4.4\cdot 10^{-5}\;.
\ee
On the other hand, an upper bound on $\alpha$ follows from the
constraint on the tensor-to-scalar ratio (\ref{cosmresults3}),
\be
\label{qalphaup}
\a<1.7\cdot 10^{-3} c_\chi^{-2}\;.
\ee
Assuming $c_\chi\sim 1$ we conclude that the model with quadratic
dispersion relation is allowed in a rather narrow parameter range
around $\alpha\sim 10^{-4}$. For the level of non-Gaussianity one
obtains,
\be
\label{qfNL}
f_{NL}=-\frac{32}{c_\chi}\bigg(\frac{10^{-4}}{\a}\bigg)^{1/2}\;,
\ee
comfortably within the allowed region (\ref{fNLresults1}). Finally,
from (\ref{qvalues1}), (\ref{cosmresults1}) we determine the energy
scale of inflation,
\be
\label{qscale}
\m\approx 5.4\cdot 10^{-3}\; \frac{c_\chi^{3/8}}{c_\T^{1/2}}\;
\bigg(\frac{\a}{10^{-4}}\bigg)^{3/16} M_0\;.
\ee
This is rather high, unlike the original model of ghost inflation
\cite{ArkaniHamed:2003uz}, which explains the tendency of our model
towards large tensor-to-scalar ratio\footnote{
After the first version of this paper has appeared on the arXiv, BICEP2 collaboration reported the discovery of the CMB B-mode
polarization, compatible with the signal from primordial tensor perturbations corresponding to the tensor-to-scalar ratio
$r=0.2^{+0.07}_{-0.05} 
\quad 
 (68\%~\mathrm{CL})\;$
\cite{Ade:2014xna}.
If confirmed, this value implies 
in the case of quadratic dispersion relation (see \eqref{qvalues3}),
\be
\a=3.8\cdot 10^{-3} c_{\chi}^{-2}\,, 
\ee
while the case of linear dispersion relation is disfavored 
(cf. \eqref{linr}). This leads to the level of non-Gaussianity
\eqref{qfNL} and the energy scale of inflation \eqref{qscale}:
$
f_{NL}\simeq-5.17\,, \quad
\m\simeq 0.01 {M_0}/{c_\T^{1/2}}\,. 
$
}.

When
\be
\label{linalphaup}
\a\ll 4.4\cdot 10^{-5}
\ee
we are in the regime of {\bf linear dispersion relation}. Now the
spectral index receives an additional contribution related to the time
variation of $\delta$, so that $\delta^2$ cannot be precisely
determined any longer. Still, without fine-tuning (\ref{linvalues2})
implies 
\be
\label{lindelta}
\delta^2\lesssim 6.6\cdot 10^{-3}\;.
\ee 
Then 
\be
\label{linr}
r= 0.026\;\bigg(\frac{\delta^2}{6.6\cdot 10^{-3}}\bigg)^{3/2}
\ee
is below the Planck sensitivity but may be accessible to future
missions \cite{Baumann:2008aq}. The predicted non-Gaussianity is
\be
\label{linfNL}
f_{NL}=-39\;\bigg(\frac{6.6\cdot 10^{-3}}{\delta^2}\bigg)\;,
\ee
again within the Planck constraint (\ref{fNLresults2}). For the scale
of inflation we obtain, 
\be
\label{linscale}
\m\approx \frac{6.4\cdot 10^{-3}}{c_\T^{1/2}}\;
\bigg(\frac{\delta^2}{6.6\cdot 10^{-3}}\bigg)^{3/8} M_0\;. 
\ee

Last but not least, we have to check that the scale of inflation is below
the theory cutoff (\ref{cutoff}), so that the model stays within the
validity of effective field theory. By substituting the numbers in
(\ref{qscale}) we find that for the case of quadratic dispersion
relation this requirement is (marginally) fulfilled. On the other
hand, for the linear dispersion relation the validity of the EFT
requires 
\be
\frac{4\cdot 10^{-5}}{c_\T}
\bigg(\frac{\delta^2}{6.6\cdot 10^{-3}}\bigg)^{3/4}\lesssim\a\;,
\ee
which is at tension with (\ref{linalphaup}). Thus we are forced to
conclude that in the kinetically driven limit the regime of purely
linear dispersion dominance is incompatible with the EFT description. 
Rather, at $\a\sim 5\cdot
10^{-5}$ both terms in the dispersion relation (\ref{mless}) are
important and the whole equation following from (\ref{gheff}) must be
solved to find the evolution of the fluctuation modes. This requires
numerical integration, which is beyond the scope of the present
paper. One expects that the results for the observables will
interpolate between the formulas (\ref{qvalues}) and
(\ref{linvalues}). The shape of the bispectrum will be a mixture
of the two shapes plotted on Fig.~\ref{Fig:1}. 

It is worth emphasizing that the problem with the inflation scale hitting
the cutoff 
appears only in the kinetically driven limit and is easily avoided by
adding the inflaton potential.

\section{Conclusions}
\label{sec:discussion}

In this paper we presented a setup that provides a partial
UV-completion (``UV-extension") of the
well-known ghost inflation model up to a scale which can be almost as high
as the Planck mass. This is achieved by coupling the inflaton to the
Lorentz-violating sector described by the Einstein-aether theory or
its khronometric version. The cutoff of our model coincides with the
cutoff of the Einstein-aether sector and is unrelated to the scale of
the ``ghost condensate'' --- the dynamically developed expectation
value for the time derivative of the inflaton. In the khronometric
version the construction can be potentially completed even further
in the framework of the Ho\v rava gravity. Curiously, the
UV-extension occurs without restoration of the broken Lorentz
symmetry, cf. \cite{Endlich:2013vfa}. 

We have studied the inflationary evolution of the Universe and the
generation of primordial perturbations in the model. The latter are
described by an effective theory containing a single degree of
freedom and governed by the same effective action 
as in the ghost inflation. 
The novelty of our model compared to the previous works on ghost
inflation is that it allows to go beyond the study of small
perturbations and incorporates in a unified framework the background
dynamics. We have found that the ghost condensate
gives positive contribution into
the energy budget of the Universe.  
This makes inflation possible even in the absence of any potential for
the inflaton --- the regime that we called kinetically driven
inflation. In principle, it is straightforward to incorporate in our
model the graceful exit from inflation and reheating which
proceeds through vanishing\footnote{At least up to the tiny present value
  of the dark energy density. We do not consider a fine-tuned
  situation when after inflation the ghost condensate stays large and
  its positive energy is cancelled by a negative cosmological
  constant.} 
of the inflaton potential and the ghost
condensate. This would be impossible in the original formulation of
the ghost inflation as the low-energy EFT because its cutoff goes to zero
in this limit. 
   
On the phenomenological side, we have calculated the characteristics
of the power spectrum and bispectrum predicted by the
model. Specifically, in the kinetically driven regime the model
predicts 
a rather high tensor-to-scalar ratio and is already
constrained by the bounds from the Planck mission. 
The non-Gaussianity is predicted to be close to the equilateral type with the amplitude ranging from $f_{NL} \sim -50$ for $r\sim 0.02$ to  $f_{NL} \sim -5$ for $r\sim 0.2$. 
This is still well within the limits set by
Planck. Optimistically, one can expect it to be probed by planned
surveys \cite{Baumann:2008aq,Amendola:2012ys,Andre:2013nfa}.

There are two directions in which our work can be extended. First,
additional insight will be gained from the study of higher
statistics. In the general EFT formulation of ghost inflation the
trispectrum depends on an additional free coupling constant standing
in front of the quartic interaction of inflaton perturbations 
$(\d_i\pi)^2 (\d_j\pi)^2$ \cite{Izumi:2010wm}. On the other hand, in
our model this coupling is unambiguously determined by the specific form
of the UV-extended theory. Repeating the analysis presented in the
beginning of Sec.~\ref{sec:bispectra} it is
straightforward to find the leading quartic interaction
\[
S_{[\pi]}^{(4)}=\int d^4x\;\frac{1}{8\m^4c_\T^6 a}\,(\d_i\pi)^2 (\d_j\pi)^2\;.
\]
Clearly, the quartic coupling is related the cubic one, see
Eq.~(\ref{Hint}), and thus the shape of the trispectrum will be
uniquely predicted.

Second, it is interesting to investigate if the (partial) UV-completion along
the lines presented in this paper can be found for other
Lorentz-violating models of modified gravity. From the theory
viewpoint, the case of Lorentz-violating massive gravity  
 \cite{Dubovsky:2004sg} deserves particular attention as its
 completion above the scale of strong coupling associated to the
 graviton mass could be considered as the analog of the Higgs
 mechanism in massive Yang--Mills theory.

\section*{Acknowledgments}
We are grateful to Diego Blas, Valery Rubakov and Adam Solomon 
for useful discussions. We thank Shinji Mukohyama for valuable comments
on the draft. 
This work was supported in part by 
the Grants of the President of Russian Federation NS-2835.2014.2 and
MK-1754.2013.2~(M.I.), the RFBR grants
14-02-31435~(M.I.), 14-02-00894~(M.I.)
and by the Dynasty Foundation~(M.I.). 

\appendix 

\section{Inflaton--metric mixing}
\label{app:A}

In this Appendix we show that for $\a\ll 1$ the mixing
between the inflaton and the metric can be neglected in the analysis of
the primordial perturbations. 
Besides the terms included in (\ref{S2FRW}), the $\T$-action
(\ref{inflAct}) contains the following terms mixing the
inflaton and the metric,\footnote{We note that $S_{[\T]}$ does not
  produce any mixing between $h_{\m\n}$ and $v_\m$ in addition to that
present in $S_{[u]}$.}
\be
\label{Lhpi2}
{\cal
  L}_{h\pi}=-\frac{(1+\vk)\dot{\bar\T}}{2}\, h^{00}\dot\pi
+\big((1+\vk){\dot{\bar\T}}-\m^2\big)\bigg(\frac{1}{2}h^i_i\dot\pi-
h^{0i}\d_i\pi\bigg)\;,
\ee
where the indices on the metric components have been raised using the
background metric $\bar g^{\m\n}$. For the sake of the argument, 
in deriving (\ref{Lhpi2}) 
 we neglected the dependence of the
functions $\vk$, $\m$, $V$ on $\T$. We see that the mixing term
contains only one derivative as opposed to the two-derivative kinetic
terms for $h_{\m\n}$ and $\pi$ and
thus is irrelevant above certain frequency $\omega_{mix}$
and momentum $k_{mix}$ \cite{Cheung:2007st}. We need to verify that
$\omega_{mix}$ is smaller than the Hubble parameter $H$, so that the
mixing stays negligible when the mode freezes out. 

As explained in the main text, the behavior of the $\pi$-modes is
qualitatively different depending on whether the physical momentum
$k/a$ is larger or smaller than $k_c$ given by (\ref{kc}). For
$k/a>k_c$ the
frequency of the mode is of the same order as the momentum,
$\omega\sim k/a$ and the mixing is estimated as,
\be
{\cal L}_{[h\pi]} \sim -\frac{\m^2}{2}h\d\pi-\m^2\delta^2 h\d\pi\;,
\ee
where we have used the background value (\ref{Tdot1}) for
${\dot{\bar\T}}$. Taking into account that $\omega>k_c$ we see that
the first term is smaller than the kinetic terms for $h_{\m\n}$ and
$\pi$ by a factor $\sqrt\a$, while the second term is further
suppressed by $\delta^2$.

At $k/a<k_c$ the dispersion relation for $\pi$ is given by (\ref{mless}) and
implies $\omega\ll k$. Then in the equation of motion for the metric
perturbations we can neglect the terms with time derivatives.
This yields schematically,
\be
M_0^2k^2 h\sim \m^2\dot\pi~~\Longrightarrow
~{\cal L}_{[h\pi]}\sim \frac{\m^4}{k^2M_0^2}\dot\pi^2\;,
\ee 
where we have omitted the contribution with the spatial derivative of
$\pi$ as it is always subdominant. 
For the modes at freeze-out, $\omega\sim H$,
\be
\label{mixlast}
{\cal L}_{[h\pi]}\sim \delta^2\frac{\m^4}{M_0^2H^2}\dot\pi^2
~~~\text{or}~~~ 
\sqrt\a\frac{\m^2}{M_0 H}\dot\pi^2
\;,
\ee
depending on whether the dispersion relation (\ref{mless}) is
dominated by the first or the second term. The Friedmann equation
(\ref{Friedmann}) implies\footnote{We exclude the fine-tuned 
 situation when the
  kinetic and potential energy of the inflaton nearly cancel each
  other.} $H\gtrsim\m^2/M_0$. Thus the contribution
(\ref{mixlast}) is in both cases 
parametrically suppressed compared to the first term
in the $\pi$-Lagrangian (\ref{gheff}). We conclude that the mixing
between $\pi$ and the metric can be consistently neglected.

\section{Scalar perturbations:  the uniform inflaton gauge}
\label{App1}

To verify the frozen-metric approximation adopted in the main text
we calculate the effective quadratic action for scalar
perturbations allowing the metric to be dynamical. For concreteness,
we concentrate on the khronometric version of the theory and work in
the gauge comoving with the khronon,
where $\sigma=t$. As we saw in Sec.~\ref{sec:lin}, at low momenta
the perturbations of the inflaton are tied to those of the khronon, see
Eq.~(\ref{tracer}), implying that surfaces of constant khronon are
also surfaces of constant inflaton. Thus in the chosen gauge the
inflaton is uniform on the constant-time slices, $\T=\bar\T(t)$. 

Writing the metric in the ADM form,
\be
ds^2=N^2 dt^2-\gamma_{ij}(dx^i+N^idt)(dx^j+N^jdt)\;,
\ee
we obtain the action
\be
\begin{split}
S\equiv S_{[EH]}+S_{[u]}+S_{[\T]}=
\int d^4x N\sqrt{\gamma}\bigg[&\frac{M_0^2}{2}\bigg((1-\b)K_{ij}K^{ij}
-(1+\l)K^2+{\cal R}+\a \frac{(\d_iN)^2}{N^2}\bigg)\\
&+\frac{(1+\vk){\dot{\bar{\T}}}^2}{N^2}
-\frac{\m^2\dot{\bar\T}}{N}-V\bigg]\;,
\end{split}
\ee
where
\[
K_{ij}=\frac{\dot \gamma_{ij}-\nabla_iN_j-\nabla_j N_i}{2N}
\]
is the extrinsic curvature of constant-time slices, $K$ is its trace,
${\cal R}$ is the intrinsic curvature constructed from the 3d-metric
$\gamma_{ij}$; the indices are raised and lowered using $\gamma_{ij}$ and its
inverse. Next, we specify to the sector of scalar perturbations,
\[
N=1+\phi~,~~~~N^i=\d_i\psi~,~~~~\gamma_{ij}=a^2\e^{2\zeta}\delta_{ij}\;,
\]
and expand the action up to quadratic order,
\be
\label{S2com}
\begin{split}
S^{(2)}=\int d^4x\bigg[&-a^3V\phi^2+aM_0^2\a\frac{(\d_i\phi)^2}{2}
+6a^2\dot aM_c^2\phi\dot\zeta-2aM_0^2\phi\D\zeta
-2\dot a M_c^2\phi\D\psi\\
&-M_0^2(\l+\b)\frac{(\D\psi)^2}{2a}
+2aM_c^2\dot\zeta\D\psi-3a^3M_c^2\dot\zeta^2
+aM_0^2(\d_i\zeta)^2\bigg]\;,
\end{split} 
\ee
where $M_c^2=M_0^2(1+\b/2+3\l/2)$. In deriving this expression we made
use of the background equations of motion (\ref{Friedmann}),
(\ref{eomPhi2}). 
Variation of (\ref{S2com}) with respect to $\phi$, $\psi$ produces the
constraints, 
\bseq
\begin{align}
-2a^3V\phi-aM_0^2\a\D\phi+6a^2\dot a M_c^2\dot\zeta-2aM_0^2\D\zeta-
2\dot a M_c^2\D\psi=0\;,\\
-M_0^2(\l+\b)\frac{\D\psi}{a}-2\dot a M_c^2\phi
+2a M_c^2\dot\zeta=0\;.
\end{align}
\eseq
Solving these equations with respect to $\phi$ and $\psi$,
substituting the result back into (\ref{S2com}) and retaining only
terms of order up to $O(\a)$ one obtains,
\be
\label{S2com1}
S^{(2)}=\int d^4x\, \frac{a^3(1+\vk)}{2}\,\frac{{\dot{\bar\T}}^2}{H^2}
\bigg[\bigg(1-\frac{(\l+\b)(1+\vk){\dot{\bar\T}}^2}{4M_0^2H^2}\bigg)
\,\dot\zeta^2-\tilde\delta^2\frac{(\d_i\zeta)^2}{a^2}
-\frac{(\l+\b)M_0^2}{(1+\vk){\dot{\bar\T}}^2}\,\frac{(\D\zeta)^2}{a^4}\bigg]
\ee
with
\be
\label{deltatilde}
\tilde\delta^2=-\frac{2M_0^2\dot H}{(1+\vk)\dot{\bar\T}}-\frac{\l+\b}{2}
=\delta^2-\frac{\l+\b}{2}\;,
\ee
where $\delta$ is defined in (\ref{delta0}). Recalling the expression for
the inflaton background (\ref{Tdot1}) one
observes that upon the substitution (\ref{zeta}) the action 
coincides with the formula (\ref{gheff}) obtained in the main text
up to negligible corrections in
the time-derivative term and the replacement
$\delta^2\mapsto\tilde\delta^2$. Let us show that the latter difference is
never important. Indeed, if $\delta^2\lesssim \l+\b$, then $\delta^2$
also satisfies the inequality (\ref{quadrcond}) and the dispersion
relation is dominated by the quadratic contribution. This implies that
the second term in (\ref{S2com1}) can be neglected altogether. On the
other hand, if $\delta^2$ satisfies (\ref{linearcond}) we have 
$\delta^2\gg H/k_c\sim\sqrt{\a}\gg \l+\b$, and the difference between
$\tilde\delta^2$ and $\delta^2$ is subleading.



\begin{thebibliography}{99}

\bibitem{Horava:2009uw} 
  P.~Horava,
  Phys.\ Rev.\ D {\bf 79}, 084008 (2009)
  [arXiv:0901.3775 [hep-th]].

\bibitem{Blas:2009qj} 
  D.~Blas, O.~Pujolas and S.~Sibiryakov,
  Phys.\ Rev.\ Lett.\  {\bf 104}, 181302 (2010)
  [arXiv:0909.3525 [hep-th]].

\bibitem{Blas:2010hb} 
  D.~Blas, O.~Pujolas and S.~Sibiryakov,
  JHEP {\bf 1104}, 018 (2011)
  [arXiv:1007.3503 [hep-th]].

\bibitem{Jacobson:2000xp}
  T.~Jacobson and D.~Mattingly,
  Phys.\ Rev.\  D {\bf 64}, 024028 (2001)
  [arXiv:gr-qc/0007031].

\bibitem{Jacobson:2008aj}
  T.~Jacobson,
  PoS {\bf QG-PH}, 020 (2007)
  [arXiv:0801.1547 [gr-qc]].
  
\bibitem{Jacobson:2010mx} 
  T.~Jacobson,
  Phys.\ Rev.\ D {\bf 81}, 101502 (2010)
  [Erratum-ibid.\ D {\bf 82}, 129901 (2010)]
  [arXiv:1001.4823 [hep-th]].
  
\bibitem{Jacobson:2013xta} 
  T.~Jacobson,
  ``Undoing the twist: the Ho\v{r}ava limit of Einstein-aether,''
  arXiv:1310.5115 [gr-qc].

\bibitem{Elliott:2005va} 
  J.~W.~Elliott, G.~D.~Moore and H.~Stoica,
  JHEP {\bf 0508}, 066 (2005)
  [hep-ph/0505211].

\bibitem{Blas:2011zd} 
  D.~Blas and H.~Sanctuary,
  Phys.\ Rev.\ D {\bf 84}, 064004 (2011)
  [arXiv:1105.5149 [gr-qc]].

\bibitem{Audren:2013dwa} 
  B.~Audren, D.~Blas, J.~Lesgourgues and S.~Sibiryakov,
  JCAP {\bf 1308}, 039 (2013)
  [arXiv:1305.0009 [astro-ph.CO], arXiv:1305.0009].

\bibitem{Shao:2013wga} 
  L.~Shao, R.~N.~Caballero, M.~Kramer, N.~Wex, D.~J.~Champion and A.~Jessner,
  Class.\ Quant.\ Grav.\  {\bf 30}, 165019 (2013)
  [arXiv:1307.2552 [gr-qc]].

\bibitem{Yagi:2013qpa} 
  K.~Yagi, D.~Blas, N.~Yunes and E.~Barausse,
  ``Strong Binary Pulsar Constraints on Lorentz Violation in Gravity,''
  arXiv:1307.6219 [gr-qc].

\bibitem{Yagi:2013ava} 
  K.~Yagi, D.~Blas, E.~Barausse and N.~Yunes,
  ``Constraints on Einstein-\AE ther theory and Horava gravity from binary pulsar observations,''
  arXiv:1311.7144 [gr-qc].

\bibitem{GrootNibbelink:2004za} 
  S.~Groot Nibbelink and M.~Pospelov,
  Phys.\ Rev.\ Lett.\  {\bf 94}, 081601 (2005)
  [hep-ph/0404271].

\bibitem{Pospelov:2010mp} 
  M.~Pospelov and Y.~Shang,
  Phys.\ Rev.\ D {\bf 85}, 105001 (2012)
  [arXiv:1010.5249 [hep-th]].

\bibitem{Pujolas:2011sk} 
  O.~Pujolas and S.~Sibiryakov,
  JHEP {\bf 1201}, 062 (2012)
  [arXiv:1109.4495 [hep-th]].

\bibitem{Bednik:2013nxa} 
  G.~Bednik, O.~Pujol\`as and S.~Sibiryakov,
  JHEP {\bf 1311}, 064 (2013)
  [arXiv:1305.0011 [hep-th]].

\bibitem{ArkaniHamed:2003uz} 
  N.~Arkani-Hamed, P.~Creminelli, S.~Mukohyama and M.~Zaldarriaga,
  JCAP {\bf 0404}, 001 (2004)
  [hep-th/0312100].

\bibitem{Senatore:2004rj} 
  L.~Senatore,
  Phys.\ Rev.\ D {\bf 71}, 043512 (2005)
  [astro-ph/0406187].

\bibitem{Ade:2013ydc} 
  P.~A.~R.~Ade {\it et al.}  [Planck Collaboration],
  ``Planck 2013 Results. XXIV. Constraints on primordial non-Gaussianity,''
  arXiv:1303.5084 [astro-ph.CO].

 \bibitem{ArkaniHamed:2003uy} 
  N.~Arkani-Hamed, H.~-C.~Cheng, M.~A.~Luty and S.~Mukohyama,
  JHEP {\bf 0405}, 074 (2004)
  [hep-th/0312099].

\bibitem{ArkaniHamed:2005gu} 
  N.~Arkani-Hamed, H.~-C.~Cheng, M.~A.~Luty, S.~Mukohyama and T.~Wiseman,
  JHEP {\bf 0701}, 036 (2007)
  [hep-ph/0507120].

\bibitem{Endlich:2013vfa} 
  S.~Endlich, A.~Nicolis and R.~Penco,
  ``UV completion without symmetry restoration,''
  arXiv:1311.6491 [hep-th].

\bibitem{Blas:2011ni} 
  D.~Blas and S.~Sibiryakov,
  Phys.\ Rev.\ D {\bf 84}, 124043 (2011)
  [arXiv:1110.2195 [hep-th]].

\bibitem{Blas:2009ck} 
  D.~Blas, O.~Pujolas and S.~Sibiryakov,
  Phys.\ Lett.\ B {\bf 688}, 350 (2010)
  [arXiv:0912.0550 [hep-th]].

\bibitem{Withers:2009qg} 
  B.~Withers,
  Class.\ Quant.\ Grav.\  {\bf 26}, 225009 (2009)
  [arXiv:0905.2446 [gr-qc]].

\bibitem{Will:2005va} 
  C.~M.~Will,
  Living Rev.\ Rel.\  {\bf 9}, 3 (2006)
  [gr-qc/0510072].

\bibitem{Donnelly:2010cr} 
  W.~Donnelly and T.~Jacobson,
  Phys.\ Rev.\ D {\bf 82}, 064032 (2010)
  [arXiv:1007.2594 [gr-qc]].

\bibitem{Solomon:2013iza} 
  A.~R.~Solomon and J.~D.~Barrow,
  Phys.\ Rev.\ D {\bf 89}, 024001 (2014)
  [arXiv:1309.4778 [astro-ph.CO]].

\bibitem{Blas:2011en}
D.~Blas and S.~Sibiryakov,
JCAP {\bf 1107} (2011) 026
[arXiv:1104.3579 [hep-th]].

\bibitem{Creminelli:2012xb} 
  P.~Creminelli, J.~Norena, M.~Pena and M.~Simonovic,
  JCAP {\bf 1211}, 032 (2012)
  [arXiv:1206.1083 [hep-th]].

\bibitem{ArmendarizPicon:2010rs} 
  C.~Armendariz-Picon, N.~F.~Sierra and J.~Garriga,
  JCAP {\bf 1007}, 010 (2010)
  [arXiv:1003.1283 [astro-ph.CO]].

\bibitem{Cheung:2007st} 
  C.~Cheung, P.~Creminelli, A.~L.~Fitzpatrick, J.~Kaplan and L.~Senatore,
  JHEP {\bf 0803}, 014 (2008)
  [arXiv:0709.0293 [hep-th]].

\bibitem{Bardeen:1983qw} 
  J.~M.~Bardeen, P.~J.~Steinhardt and M.~S.~Turner,
  Phys.\ Rev.\ D {\bf 28}, 679 (1983).

\bibitem{Salopek:1990jq} 
  D.~S.~Salopek and J.~R.~Bond,
  Phys.\ Rev.\ D {\bf 42}, 3936 (1990).
    
\bibitem{Maldacena:2002vr} 
  J.~M.~Maldacena,
  JHEP {\bf 0305}, 013 (2003)
  [astro-ph/0210603].

\bibitem{Babich:2004gb}
  D.~Babich, P.~Creminelli and M.~Zaldarriaga,
  JCAP {\bf 0408} (2004) 009
  [astro-ph/0405356].

\bibitem{Ade:2013zuv} 
  P.~A.~R.~Ade {\it et al.}  [Planck Collaboration],
  ``Planck 2013 results. XVI. Cosmological parameters,''
  arXiv:1303.5076 [astro-ph.CO].

\bibitem{Ade:2013uln} 
  P.~A.~R.~Ade {\it et al.}  [Planck Collaboration],
  ``Planck 2013 results. XXII. Constraints on inflation,''
  arXiv:1303.5082 [astro-ph.CO].

\bibitem{Baumann:2008aq} 
  D.~Baumann {\it et al.}  [CMBPol Study Team Collaboration],
  AIP Conf.\ Proc.\  {\bf 1141}, 10 (2009)
  [arXiv:0811.3919 [astro-ph]].

\bibitem{Amendola:2012ys} 
  L.~Amendola {\it et al.}  [Euclid Theory Working Group Collaboration],
  Living Rev.\ Rel.\  {\bf 16}, 6 (2013)
  [arXiv:1206.1225 [astro-ph.CO]].
  
\bibitem{Andre:2013nfa} 
  P.~Andr\'e, C.~Baccigalupi, A.~Banday, D.~Barbosa, B.~Barreiro, J.~Bartlett, N.~Bartolo and E.~Battistelli {\it et al.},
  ``The Polarized Radiation Imaging and Spectroscopy Mission,''
  arXiv:1310.1554 [astro-ph.CO].

\bibitem{Izumi:2010wm} 
  K.~Izumi and S.~Mukohyama,
  JCAP {\bf 1006}, 016 (2010)
  [arXiv:1004.1776 [hep-th]].

\bibitem{Dubovsky:2004sg} 
  S.~L.~Dubovsky,
  JHEP {\bf 0410}, 076 (2004)
  [hep-th/0409124].

\bibitem{Ade:2014xna} 
  P.~A.~R.~Ade {\it et al.}  [BICEP2 Collaboration],
  ``BICEP2 I: Detection Of B-mode Polarization at Degree Angular Scales,''
  arXiv:1403.3985 [astro-ph.CO].


  
  
  
  
%
%
  
  
  
  
  
  
  
  
  
%
  
    

\end{thebibliography}
\end{document}